\documentclass[aps,prl,twocolumn,preprintnumbers,superscriptaddress,10pt]{revtex4}

\usepackage{amssymb}
\usepackage{amsmath}
\usepackage{amsfonts}
\usepackage{graphicx}
\usepackage{subfigure}
\usepackage{color}
\usepackage{dcolumn}
\usepackage{hyperref}
\newcommand{\be}{\begin{equation}}
\newcommand{\ee}{\end{equation}}
\newcommand{\bea}{\begin{eqnarray}}
\newcommand{\eea}{\end{eqnarray}}
\newcommand{\nn}{\nonumber}
\newcommand{\rd}{\partial}

 \def\cN{{\cal N}}

\def\a{\alpha}      
\def\b{\beta}       
\def\g{\gamma}

\def\k{\kappa}

\def\o{\omega}

\def\s{\sigma}  
\def\t{\tau}

\newcommand{\abs}[1]{{\left| {#1} \right|}}
\newcommand{\prt}[1]{{\left( {#1} \right)}}
\newcommand{\eq}[1]{(\ref{#1})}
\newcommand{\la}[1]{\label{#1}}

\begin{document}

\title{Universal Properties of the Langevin Diffusion Coefficients }

\author{Dimitrios Giataganas}
\affiliation{Physics Division, National Technical University of Athens, 15780 Zografou Campus, Athens, Greece}
\author{Hesam Soltanpanahi}
\affiliation{Institute of Physics, Jagiellonian University, Reymonta 4, 30-059 Krakow, Poland}
\affiliation{National Institute for Theoretical Physics, School of Physics and Centre for Theoretical Physics, University of the Witwatersrand, Wits 2050, South Africa}

\preprint{WITS-CTP-121}

\begin{abstract}
We show that in generic isotropic holographic theories the longitudinal Langevin diffusion coefficient along the string motion is larger compared to that of the transverse direction. We argue that this is in general a universal relation and we derive the generic conditions in order to be satisfied. A way to violate the relation is to consider anisotropic gauge/gravity dualities. We give an explicit example of this violation where the noise along the transverse direction is larger than the noise occurring along the quark motion.
Moreover, we derive the effective world-sheet temperature for any generic theory and then the conditions for negative excess noise. We argue that isotropic theories can not have negative excess noise and we  additionally remark that these conditions are difficult to get satisfied, indicating positivity of the excess noise even in a large class of anisotropic holographic theories, implying a strong universal property.
\end{abstract}

\maketitle


\noindent {\bf 1. Introduction.}
Using the AdS/CFT \cite{Maldacena:1997re} correspondence we can study the dynamics of the strongly coupled quantum field theories. The rapid development and understanding of the duality has led to a large number of different gauge/gravity dualities. Therefore, it is very intriguing to find universal properties of such theories with the hope that they carry on, to some extent, in the QCD itself. A realization of such results can be applied to the quark-gluon plasma  (QGP) (see \cite{review} for a review of applications to the QGP).

An observable studied using gauge/gravity techniques is the drag force of a quark moving in the QGP \cite{Gubser:2006bz}, which can be measured from the momentum flowing to the bulk of a long open trailing string starting from a heavy quark on the boundary. The quantum fluctuations of the trailing string, are related to the momentum broadening of a heavy quark moving in the QGP, and can be analyzed independently along the direction of the motion of the quark and along the transverse plane. The stochastic motion is associated with the correlators of the fluctuations of the trailing string \cite{Gubser:2006nz,CasalderreySolana:2007qw}, and may be analyzed with the Langevin coefficients \cite{deBoer:2008gu,Son:2009vu}. The relativistic Langevin evolutions were studied in \cite{Giecold:2009cg} and in non-conformal frameworks in \cite{Gursoy:2010aa, Kiritsis:2011bw}. Since the system is out of equilibrium, the Hawking temperature of the induced string world-sheet metric $T_{ws}$ is in general different from the heat bath temperature $T$ and may be even lower than it. The relevant Langevin coefficients resembling the noise for some degrees of freedom, can be negative, implying a negative excess noise generated by the driving force \cite{Nakamura:2013yqa}. This is of particular interest since certain quantum conductors can have absent or negative excess noise \cite{NEN}.

In this paper we analyze in a completely generic framework the relativistic Langevin coefficients of a heavy quark using the gauge/gravity duality techniques. We find that a relation, noticed also to hold in \cite{Gursoy:2010aa} for particular class of backgrounds \footnote{In \cite{Gursoy:2010aa} the form of the metric used was $ds^2=b\prt{u}^2\prt{f\prt{u}^{-1} du^2-f\prt{u}dt^2+dx_i dx^i}$, with its exact details written in the paper.}, namely an inequality between the transverse and longitudinal to the motion Langevin diffusion coefficients $\k_\parallel>\k_\perp$, holds for generic isotropic theories including ones with certain Lifshitz scalings. We derive the conditions to hold in order to satisfy this universal relation, and we express them in a compact form in terms of the metric elements of the gravity dual theory. They turn out to be satisfied by most of known theories, and that is why we argue that the inequality of the Langevin coefficients is in general universal, while we explicitly find the range of the universality.

A way to violate the universal relation between the diffusion coefficients is to consider a moving quark in anisotropic theory. In some sense this is similar to the violation of the shear viscosity over entropy density bound \cite{univ1,Iqbal:2008by} analyzed in \cite{Rebhan:2011vd}. We give a concrete example by calculating the Langevin coefficients in the space-dependent axion anisotropic theory \cite{Mateos:2011ix}.

We also derive the generic formula of the world-sheet temperature $T_{ws}$ and we note that in most theories $T_{ws}<T$ having holographic refrigerator systems, while in the anisotropic theories the inequality may change \footnote{A detailed study on that will appear in \cite{future}.}.

Another part of our motivation is to derive the condition for negative excess noise in a quark moving in a plasma. We find this condition in a compact form, expressed in terms of the time and the spatial metric elements. We note that this condition is stricter than the previous ones and is more difficult to obtain theories with negative excess noise, implying again a universal behavior. We argue that the only way to have negative excess noise is for a moving quark in anisotropic systems. Although we find that in the axion deformed anisotropic theory \cite{Mateos:2011ix} the excess noise is positive.

\noindent {\bf 2. Setup, trailing string and drag force.}
We consider a generic background of the form
\be
ds^2=G_{00}dt^2+G_{uu}du^2+G_{ii}dx_i^2~,
\ee
which is diagonal and allows the study of anisotropic cases when at least one element of the $G_{ii}$, $i=1, 2, 3$ is not equal to the other two. All components are functions of radial coordinate $u$, the boundary of the space is taken at $u\rightarrow 0$ and the element $G_{00}$ contains the blackening factor with the black hole horizon position.

The trailing string corresponding to a quark moving on the boundary along
the chosen direction $x_p$, $p=1, 2, 3,$ with a constant velocity $v$, is characterized by the following parametrization $t=\t,~u=\sigma,~x_p=v~t+\xi(u)~,$ 
and localized in the rest of dimensions, which results to the induced world-sheet metric
\begin{eqnarray}
g_{\a\b}=\left(\begin{array}{cc}
G_{00}+v^2\,G_{pp}&G_{pp}\,v\,\xi'\\
G_{pp}\,v\,\xi'&G_{uu}+\xi'^2\,G_{pp}
\end{array}\right)\la{ind1}~.
\end{eqnarray}
Taking the Nambu-Goto (NG) action we find that the momentum $\Pi^p_u$  flowing from the boundary to the bulk, which is constant of motion, is related to  $\xi'$ by
\be\la{xi11}
\xi'^2=-G_{uu} C^2\,\frac{G_{00}+G_{pp}\,v^2} {G_{00}\,G_{pp}\prt{C^{2}+G_{00}\,G_{pp}}}~,
\ee
where $C=2\,\pi\,\alpha'\,\Pi^p_u$. There is a critical point at which both numerator and denominator change their sign. This point $u_0$ is defined by
\be
G_{00}(u_0)=- G_{pp}(u_0)\,v^2~,\label{ucond}
\ee
where we have supposed $G_{uu}(u_0)\neq 0$.
The drag force can be written in the compact form $F_{drag,x_p}=
-v\, G_{pp}(u_0)\prt{2\pi\alpha'}^{-1}~.$

\noindent {\bf 2-1. World-sheet temperature.}
The world-sheet horizon obtained by $g_{\tau\tau}(\s_h)=0$ from \eq{ind1} and the critical point $u_0$ are the same, and they are both obtained by solving the eq. \eqref{ucond}. In order to find the effective temperature of the world-sheet horizon we diagonalize the world-sheet metric by introducing a new coordinate as
\be
{\tau}\rightarrow \t+A(u),~~~~A'(u)=\,-\frac{v \xi' G_{pp}}{G_{00}+v^2\,G_{pp}},
\ee
or equivalently $d\t\rightarrow d\t-g_{\t\s}/g_{\t\t}~d\s$.
The metric becomes diagonal with the following components 
\bea\label{diagonalize-metric1}
h_{\s\s}
=\frac{G_{00}G_{uu}G_{pp}}{G_{00}G_{pp}+C^2}\,,\quad
h_{{\tau}{\tau}}=G_{00}+v^2\,G_{pp}\,,
\eea
and the corresponding effective temperature can be found from the coordinate singularity, giving
\bea
\la{tws1}
T_{ws}^2=
\frac{1}{16\pi^2}\bigg|\frac{1}{G_{00} G_{uu}}\prt{G_{00}\,G_{pp}}' \prt{\frac{G_{00}}{G_{pp}}}'\bigg|\Bigg|_{u=u_0}~,
\eea
 expressed solely in terms of the background metric
elements.

\noindent {\bf 3. Fluctuations of the Trailing String.}
We add fluctuations in classical trailing string solutions as in \cite{Gubser:2006nz}, in order to study the stochastic motion of the quark. We choose the static gauge 
and consider fluctuations of the form $\delta x(\tau, \sigma)$ along longitudinal and transverse coordinates. The induced metric on the world-sheet is given by $\tilde{g}_{\a\b}=g_{\a\b}+\delta g_{\a\b}~,$
resulting in the following NG action for fluctuations around the solution \eq{xi11} to quadratic order
\bea
S_2&&=-\frac{1}{2\pi\alpha'}\int d\t d\s \sqrt{-g}\,\frac{g^{\a\b}}{2}\times\nn\\
&&\left[N(u)\rd_\a \delta x_p\,\rd_\b \delta x_p+\sum_{i\neq p}{G_{ii}}\rd_\a \delta x_i\,\rd_\b \delta x_i\right]~,\nn
\eea
where we have used the 'on-shell' expression for the world-sheet  determinant
\be
g=G_{00}\,G_{uu}\,G_{pp}\,\frac{G_{00}+G_{pp}\,v^2}{G_{00}\,G_{pp}+C^2}~,
\ee
and
\be
N(u)={\frac{G_{00}\, G_{pp}+C^2}{G_{00}+G_{pp}\,v^2}}.
\ee
The linear terms in fluctuations form a total derivative and can be neglected with the particular boundary conditions imposed.
We may rewrite the above action in terms of the coordinates that diagonalize the world-sheet metric with the metric elements \eqref{diagonalize-metric1}. It takes the following form
\bea\la{s22}
S_2&&=-\frac{1}{2\pi\alpha'}\int d\tau d\s \,\frac{H^{\a\b}}{2}\times\\
&&\left[N(u)\,\rd_\a \delta x_p\,\rd_\b \delta x_p+\sum_{i\neq p}{G_{ii}}\rd_\a \delta x_i\,\rd_\b \delta x_i\right]~,\nn
\eea
where $H^{\a\b}=\sqrt{-h}{h}^{\a\b},$ and  $h^{\a\b}$ is inverse of the diagonalized induced world-sheet metric \eqref{diagonalize-metric1}.
The equations of motion for fluctuations are derived from the action \eq{s22} and are
\bea
\rd_\a\left(H^{\a\b}N \rd_\b \delta x_{p}\right)=0,~~\rd_\a\left( H^{\a\b}G_{ij}\rd_\b \delta x_j\right)=0~.
\eea
A way to compute the Langevin correlation functions from the
fluctuation equations and extract the diffusion constants and the spectral densities is to use the harmonic ansatz $\delta x_i(\tau,\s)=e^{i\omega\tau}\delta x_i(\omega,\s)$. Then the equations of motion become
\bea
&&\rd_\s\left(\frac{N\,g_{\t\t}}{\sqrt{-g}}  \, \rd_\s \delta x_{p}\right)+\omega^2\frac{N\sqrt{-g}}{g_{\t\t}} \delta x_{p}=0~,\nn\\
&&\rd_\s\left(\frac{G_{ij}\,g_{\t\t}}{\sqrt{-g}}  \, \rd_\s \delta x_{i}\right)+\omega^2\frac{G_{ij}\sqrt{-g}}{g_{\t\t}} \delta x_{i}=0~.\nn
\eea
To write the above equations we have used the fact that the $h_{\s\s}$ world-sheet metric element may be written as $g/g_{\t\t}$, which from \eq{diagonalize-metric1} results in $h=g$.

\noindent {\bf 3-1. Langevin Coefficients.}
Our metric form is a generic diagonal and is chosen in such a way that the metric elements depended only on the radial coordinate of the space.
The Langevin analysis for the fluctuations we have introduced leads to two types of independent retarded correlators $G_{R}^\parallel\prt{\omega}$ and $G_R^\perp\prt{\omega}$  \cite{Gubser:2006nz, Gursoy:2010aa} for the longitudinal and transverse fluctuations.
A direct way to calculate the diffusion coefficients is by using the membrane paradigm \cite{Iqbal:2008by} for the world-sheet action. Using the effective action \eq{s22}, we obtain the transport coefficients associated with the massless fluctuations, from their coupling to the effective action evaluated at the horizon. In our generic theory where anisotropies are allowed along the different directions, this can be done consistently for motion of the quark along each direction. For each quark direction there is a unique horizon that makes the application of membrane paradigm possible. In other words we are taking advantage of the fact that the resulting two-dimensional world-sheet metric has some universal properties, irrelevant of how complicated the initial background metric is.
From \cite{CasalderreySolana:2007qw, Gubser:2006nz, Giecold:2009cg} we get the
imaginary part of the retarded correlator related to the symmetrized correlator by $G\prt{\omega}=-\coth\prt{\omega/\prt{2 T_{ws}}} \text{Im}G_R\prt{\o} $
where the world-sheet temperature dependence enters.  The form of the diffusion constant follows then from the Langevin boundary analysis giving
$\k_{a}=\lim_{\o\rightarrow 0} G^a\prt{\omega}=-2\,T_{ws}\,\lim_{\omega\rightarrow 0}\,{\rm Im}G^a_{R}(\omega)/\omega~.$ The index $a$ refers to longitudinal and transverse fluctuations $\prt{\k_\parallel,~\k_\perp}$.
Here we use the fact that a fluctuation $\phi$ in the bulk of a generic theory leads to an action of the form
\be
S_2=-\frac{1}{2}\int dx du \sqrt{-g} q \prt{u}g^{ \a \b}\partial_\a \phi \partial_\b \phi
\ee
where the relevant transport coefficients associated with the retarded Green functions can be read directly from the action, with $q$ being the crucial input, since in two dimensions the rest of the induced metric dependence get canceled.
The identification should be done with the action \eq{s22} which gives the generic formulas for the transverse fluctuations
\bea
q_\perp&=&\frac{1}{2\pi\alpha'}\,G_{kk}\bigg|_{u=u_0},
\eea
where the index $k$ denotes a particular transverse to motion direction and no summation is taken. For the longitudinal fluctuations
\bea\la{qpar}
q_\parallel&=&
\frac{1}{2\pi\alpha'}\,\frac{\left(G_{00}G_{pp}\right)'} {G_{pp}\,\left(\frac{G_{00}}{G_{pp}}\right)'}\Bigg|_{u=u_0}~,
\eea
where we have employed L' Hospital's rule since our functions are continuous. Therefore the Langevin coefficients read
\bea
\k_\perp&=&\frac{1}{\pi\alpha'}\,G_{kk}\bigg|_{u=u_0} T_{ws},\label{mpa1}\\
\k_\parallel&=&\pm\frac{16\,\pi}{\alpha'}\,
\frac{{\abs{G_{00}}\,G_{uu}}}{G_{pp}\, {\left(\frac{G_{00}}{G_{pp}}\right)'\abs{\left(\frac{G_{00}}{G_{pp}}\right)'}}} \Bigg|_{u=u_0}T_{ws}^3~,~~{} \label{mpa2}
\eea
expressed completely in the initial metric elements, where the $T_{ws}$ is given by \eq{tws1} and the sign $\pm$ follows the sign of $\left(G_{00}G_{pp}\right)'$.
Their ratio can be written as
\be\la{ratio1}
\frac{\k_\parallel}{\k_\perp}=\pm16 \pi^2\frac{{\abs{G_{00}}\,G_{uu}}}{G_{kk}G_{pp}\, {\left(\frac{G_{00}}{G_{pp}}\right)'\abs{\left(\frac{G_{00}}{G_{pp}}\right)'}}} \Bigg|_{u=u_0}T_{ws}^2~.
\ee

\noindent {\bf 3-2. $\k_\parallel>\k_\perp$ for generic isotropic theories.} \label{klktiso}
For particular gauge/gravity dualities the inequality $\k_\parallel>\k_\perp$ has been noticed to hold \cite{Gursoy:2010aa, Nakamura:2013yqa}. In this section we examine the generic conditions for the inequality to be satisfied. We claim that it is a universal relation, since to violate these conditions, very particular backgrounds need to be considered. In a sense this arguing is similar to the shear viscosity over entropy bound. The universality of the ratio or the lower value bound are extensively violated only when very particular backgrounds are considered \cite{Rebhan:2011vd,Erdmenger:2010xm}.

Let us express the Langevin coefficients ratio in terms of the metric elements exclusively. The equation \eq{ratio1} takes the compact form
\be\la{ratio2}
\frac{\k_\parallel}{\k_\perp}=\frac{\prt{G_{00}G_{pp}}'}{G_{kk}G_{pp}\, {\left(\frac{G_{00}}{G_{pp}}\right)'}} \Bigg|_{u=u_0}
\ee
which can be brought for our purposes to
\be\la{ratio3}
\frac{\k_\parallel}{\k_\perp}=\frac{G_{pp}}{G_{kk}}\prt{1+ \frac{2 G_{00}G_{pp}'}{G_{00}'G_{pp}-G_{00}G_{pp}'} }\Bigg|_{u=u_0}~.
\ee
For an isotropic background, $G_{pp}=G_{kk}$. We assume that the boundary of the space is at $u=0$, where the elements $|G_{00}|$ and $G_{pp}$ take their maximum values and as the radial coordinate of the space increase, these are decreasing functions \footnote{The boundary position of the space does not affect the outcome of this discussion.}. The numerator $G_{00} G_{pp}'>0$ takes positive values outside the horizon of the black hole of the initial background metric. The denominator reduces to a problem of finding the sign of the ratio $r_1=\prt{G_{00}/G_{pp}}'$. Consider the cases where $G_{00}=-G_{pp} G_{bh}$, where $G_{bh}\propto 1-u^m/u_h^m$, with $m>0$, is the blackening function containing the horizon dependence, and its form is partly fixed by the assumption of the position of the boundary of the space.  The geometry of the heat bath of $Dp$-branes is a particular example that belong to this class of backgrounds. Under these conditions we get $r_1>0$ which leads
to
\be\la{universal}
\k_\parallel>\k_\perp~.
\ee
This is a universal result for this class of isotropic backgrounds, while the equality is satisfied for $v\rightarrow 0$.

For the more general case, isotropic metrics with a Lifshitz scaling $G_{00}=-G_{pp} G_{bh} u^{-n}$ need to be considered. The sign of the denominator of \eq{ratio3} is specified by
$r_1= -G_{bh}'  u^{-n}+ G_{bh} n  u^{-n-1}$. If $n\geq 0 $ \footnote{ Which in the usual notation of the Lifshitz metrics would correspond to $z\geq 1$.} then the universal inequality \eq{universal} holds. When $n<0$, the background should be examined case by case, while the null energy conditions might not allow physical dual gauge theories. The case by case examination should also be done using the formula \eq{ratio2} when the black hole backgrounds have more complicated forms. For example the charged Reissner-Nordstr\"{o}m-AdS black holes satisfy the universal inequality.

Therefore, for the generic isotropic theories with the nonstrict assumptions we have made, the inequality \eq{universal} holds. Theories with Lifshitz scalings, of $n\geq 0$ satisfy always the inequality, independently of the exact details of the background.

For the anisotropic theories, the ratio \eq{ratio3}, can be violated. Even when the expression in brackets is larger than the unit, the relation between the Langevin components depend on the ratio of the metric elements along the transverse direction and the quark motion's direction. This is the way that the violation of the inequality is implemented using the space-dependent axion deformed anisotropic background \cite{Mateos:2011ix}.

\noindent {\bf 4. Violation of the Universal Inequality in the Anisotropic Theories.}\label{violation1}
The theory we use in this section is a deformed version of the $\cN=4$ sYM  with a space-dependent axion \cite{Mateos:2011ix}. The anisotropic direction is the $x_3$ direction, while the $x_1, x_2$ can be thought of as an isotropic plane.

We may consider the motion of the quark along the anisotropic direction and independently a quark motion along the transverse to anisotropic direction. For each quark direction there is a unique horizon of the induced metric and the relevant coefficients can be obtained from equations \eq{mpa1} and \eq{mpa2}.

We find for this background that $G_{11}/G_{33}$ can be less than one and therefore according to the discussion of the previous section and eq. \eq{ratio3}, the universal isotropic inequality \eq{universal} may be violated in this theory for a quark moving along the transverse direction to the anisotropy and having transverse momentum broadening along the anisotropic direction.
Indeed for a small anisotropy, using the world-sheet horizon $u_0$ found in \cite{Giataganas_aniso}, we find
\be
\frac{\k_\parallel}{\k_\perp}=
\gamma^2+\frac{a^2}{24\,\pi^2\,T^2}\,{B(\gamma)}+\mathcal{O}(\frac{a^4}{T^4})\,,
\ee
where $a$ is the parameter measuring the degree of anisotropy and $T$ is the temperature of the theory. The function $B$ defined as
\be
B(\gamma):=\left(\g-1\right) \left(4+\g+2 \g^2\right)-2 \g^2 \left(2+\g^2\right) \log\prt{1+\g^{-1}},\nn
\ee
where $\gamma:=(1-v^2)^{-1/2}$. As the anisotropic parameter $a$ increases, remaining in the small anisotropy limit, we find the inequality $\k_\parallel<\k_\perp$ for larger and larger velocities. In fact in the regime of larger anisotropies where the background is known only numerically, the inequality
\eq{universal} does not hold for the most of  range of velocities. The detailed study in the anisotropic theories can be found in \cite{future}.

\noindent {\bf 5. Negative Excess Noise in QGP.}
For a quark moving in an isotropic plasma with the properties described in section 3, it is not possible to have negative excess noise. The reason is that the inequality \eq{universal} is satisfied and the $\k_\perp$ is proportional to the transverse to the direction of motion metric element and to the $T_{ws}$, both of them positive. However, we may have negative excess noise for the longitudinal component for a quark moving in an anisotropic plasma.  For a negative excess noise we need
\be\la{nnoise}
\frac{\left(G_{00}G_{pp}\right)'} {\left(\frac{G_{00}}{G_{pp}}\right)'}\Bigg|_{u=u_0}<0~.
\ee
For certain different scalings between the metric elements, it is in principle possible to satisfy the above inequality and a negative excess noise along the quark motion in the anisotropic systems. However, we note that the condition \eq{nnoise} for negative excess noise is much stricter than the other conditions derived in this paper, and as a result it is more difficult to find holographic systems to satisfy it.  For example, in the Lifshitz-like anisotropic theory \cite{Mateos:2011ix} we find no negative Langevin coefficients.

\noindent{\bf Acknowledgments.} We would like to thank C. Hoyos, R. Janik and J. Wu for useful discussions. D.G. would like to thank the Kavli IPMU institute, the University of Tokyo, and the National Tsing-Hua University for hospitality. The research of D.G is implemented under the "ARISTEIA" action of the "operational program education and lifelong learning" and is cofunded by the European Social Fund (ESF) and National Resources. H.S. is supported by the South African National Research Foundation (NRF) and Foundation for Polish Science MPD Programme cofinanced by the European Regional Development Fund, agreement No. MPD/2009/6.

\end{document}